\documentclass{aa}  

\usepackage{graphicx}
\usepackage{txfonts}
\usepackage{color}
\usepackage{multirow}

\newcommand{\tablenotea}[1]{\parbox{12.1cm}{\indent \footnotesize{#1}}}
\newcommand{\tablenoteb}[1]{\parbox{8.8cm}{\indent \footnotesize{#1}}}
\newcommand{\cpl}{Chem. Phys. Lett.}

\newcommand{\jmst}{J. Mol. Struct.}

\newcommand{\jpcrd}{J. Phys. Chem. Ref. Data}

\newcommand{\nature}{Nature}

\newcommand{\pss}{Planet. Space Sci.}

\begin{document}

\title{Discovery of interstellar NC$_4$NH$^+$:\\ dicyanopolyynes are indeed abundant in space\thanks{Based on observations carried out with the Yebes 40m telescope (projects 19A003, 20A014, 20D023, 21A011, and 21D005). The 40m radio telescope at Yebes Observatory is operated by the Spanish Geographic Institute (IGN; Ministerio de Transportes, Movilidad y Agenda Urbana).}}

\titlerunning{Discovery of interstellar NC$_4$NH$^+$}
\authorrunning{Ag\'undez et al.}

\author{M.~Ag\'undez\inst{1}, C.~Cabezas\inst{1}, N.~Marcelino\inst{2,3}, R.~Fuentetaja\inst{1}, B.~Tercero\inst{2,3}, P.~de~Vicente\inst{3}, \and J.~Cernicharo\inst{1}}

\institute{
Instituto de F\'isica Fundamental, CSIC, Calle Serrano 123, E-28006 Madrid, Spain\\ \email{marcelino.agundez@csic.es, jose.cernicharo@csic.es} 
\and
Observatorio Astron\'omico Nacional, IGN, Calle Alfonso XII 3, E-28014 Madrid, Spain 
\and
Observatorio de Yebes, IGN, Cerro de la Palera s/n, E-19141 Yebes, Guadalajara, Spain
}

\date{Received; accepted}

 
\abstract
{The previous detection of two species related to the non polar molecule cyanogen (NCCN), its protonated form (NCCNH$^+$) and one metastable isomer (CNCN), in cold dense clouds supported the hypothesis that dicyanopolyynes are abundant in space. Here we report the first identification in space of NC$_4$NH$^+$, which is the protonated form of NC$_4$N, the second member of the series of dicyanopolyynes after NCCN. The detection was based on the observation of six harmonically related lines within the Yebes 40m line survey of \mbox{TMC-1} QUIJOTE. The six lines can be fitted to a rotational constant $B$\,=\,1293.90840\,$\pm$\,0.00060 MHz and a centrifugal distortion constant $D$\,=\,28.59\,$\pm$\,1.21 Hz. We confidently assign this series of lines to NC$_4$NH$^+$ based on high-level ab initio calculations, which supports the previous identification of HC$_5$NH$^+$ by \cite{Marcelino2020} from the observation of a series of lines with a rotational constant 2 MHz lower than that derived here. The column density of NC$_4$NH$^+$ in \mbox{TMC-1} is (1.1\,$^{+1.4}_{-0.6}$)\,$\times$\,10$^{10}$ cm$^{-2}$, which implies that NC$_4$NH$^+$ is eight times less abundant than NCCNH$^+$. The species CNCN, previously reported toward L483 and tentatively in \mbox{TMC-1}, is confirmed in this latter source. We estimate that NCCN and NC$_4$N are present in \mbox{TMC-1} with abundances a few times to one order of magnitude lower than HC$_3$N and HC$_5$N, respectively. This means that dicyanopolyynes NC$-$(CC)$_n$$-$CN are present at a lower level than the corresponding monocyanopolyynes HCC$-$(CC)$_n$$-$CN. The reactions of the radicals CN and C$_3$N with HNC arise as the most likely formation pathways to NCCN and NC$_4$N in cold dense clouds.}


\maketitle

\section{Introduction}

Dicyanopolyynes are molecules with a linear unsaturated skeleton of carbon atoms terminated at each edge by a cyano group, that is, N$\equiv$C$-$(C$\equiv$C)$_n-$C$\equiv$N. The simplest member of this family, cyanogen (NCCN), has been long known to be present in the atmosphere of Titan \citep{Kunde1981,Coustenis1991,Sylvestre2020} and has been recently identified in the coma of comet 67P/Churyumov-Gerasimenko \citep{Hanni2021}. The second member of the series, dicyanoacetylene (NC$_4$N), has been observed in solid state, although not in the gas phase, in the atmosphere of Titan \citep{Samuelson1997,Jolly2015}. It is also interesting to note that NCCP, a species chemically related to NCCN that results from the substitution of one N atom by one P atom, has been identified tentatively in the carbon star envelope IRC\,+10216 \citep{Agundez2014}.

It has been hypothesized that dicyanopolyynes could be abundant in interstellar and circumstellar clouds \citep{Kolos2000,Petrie2003}. Indeed, it is well known that monocyanopolyynes, more commonly known as cyanopolyynes, i.e., H$-$C$\equiv$C$-$(C$\equiv$C)$_n-$C$\equiv$N, are abundant in cold dense clouds and circumstellar envelopes around evolved stars, and thus it is conceivable that dicyanopolyynes are abundant as well. However, dicyanopolyynes are non polar and thus cannot be detected at radio wavelengths. In spite of that, in recent years we learnt that dicyanopolyynes are very likely abundant in interstellar clouds. The protonated form of cyanogen, NCCNH$^+$, was detected in the cold dense clouds TMC-1 and L483 \citep{Agundez2015}. Just as N$_2$H$^+$ is used to probe N$_2$ \citep{Linke1983}, the detection of NCCNH$^+$ was interpreted as an indirect evidence of the presence of NCCN, for which a large abundance, in the range (1-10)\,$\times$\,10$^{-8}$ relative to H$_2$, was inferred. A second piece of evidence in support of the presence of cyanogen in interstellar clouds was provided by the detection of isocyanogen (CNCN) in the cold dense clouds L483, TMC-1 (tentatively), and L1544 \citep{Agundez2018,Vastel2019}. This species is a polar metastable isomer of NCCN, and thus it is very likely chemically connected to cyanogen. From the detection of CNCN, \cite{Agundez2018} inferred an abundance of NCCN in the range (2.5-4.5)\,$\times$\,10$^{-9}$ relative to H$_2$.

Here we present an additional evidence in support of the existence of abundant dicyanopolyynes in the interstellar medium. We identified a series of six harmonically related lines in the QUIJOTE\footnote{Q-band Ultrasensitive Inspection Journey to the Obscure TMC-1 Environment.} line survey of TMC-1 \citep{Cernicharo2021c,Cernicharo2022b}, which are convincingly assigned to NC$_4$NH$^+$. This is the protonated form of dicyanoacetylene (NC$_4$N), and thus indirectly probes the presence of the second member of the family of dicyanopolyynes. We discuss how abundant are likely to be interstellar dicyanopolyynes in comparison with their monocyano counterparts.

\section{Observations}

The observational data used here is based on QUIJOTE \citep{Cernicharo2021c,Cernicharo2022b}, an ongoing line survey that is being carried out with the Yebes 40m telescope at the position of the cyanopolyyne peak of \mbox{TMC-1}, $\alpha_{J2000}=4^{\rm h} 41^{\rm  m} 41.9^{\rm s}$ and $\delta_{J2000}=+25^\circ 41' 27.0''$. QUIJOTE uses a 7 mm receiver covering the Q band (31.0-50.3 GHz) with horizontal and vertical polarizations. The backend is a fast Fourier transform spectrometer providing a spectral resolution of 38.15 kHz and a bandwidth of 8\,$\times$\,2.5 GHz in each polarization, which allows to cover almost completely the whole Q band. The system is described by \citet{Tercero2021}. The data presented here were acquired from November 2019 to May 2022 using the frequency-switching technique, and comprise 546 h of on-source telescope time. The intensity scale used is antenna temperature, $T_A^*$, which has an estimated uncertainty of 10~\% and can be converted to main beam brightness temperature, $T_{mb}$, by dividing by $B_{\rm eff}$/$F_{\rm eff}$, where $B_{\rm eff}$ and $F_{\rm eff}$ are the beam and forward efficiencies, respectively. For the Yebes 40m telescope in the Q band\footnote{\texttt{https://rt40m.oan.es/rt40m\_en.php}}, $F_{\rm eff}$\,=\,0.97 and $B_{\rm eff}$ can be fitted as a function of frequency as $B_{\rm eff}$\,=\,0.797\,$\exp$[$-$($\nu$(GHz)/71.1)$^2$], where the expression corresponds to values measured during 2022 that represent a slight improvement over previous values due to a better alignment of mirrors in the receiver cabin. The half power beam width (HPBW) can also be fitted as a function of frequency as HPBW($''$)\,=\,1763/$\nu$(GHz). Some spectra of \mbox{TMC-1} presented in Sec.~\ref{sec:discussion} were observed with the IRAM\,30m telescope. These observations are described in \cite{Cabezas2022a}. All data were analyzed using the GILDAS software\footnote{\texttt{http://www.iram.fr/IRAMFR/GILDAS/}}.

\section{Results}

\subsection{Identification of NC$_4$NH$^+$}

\begin{figure}
\centering
\includegraphics[angle=0,width=\columnwidth]{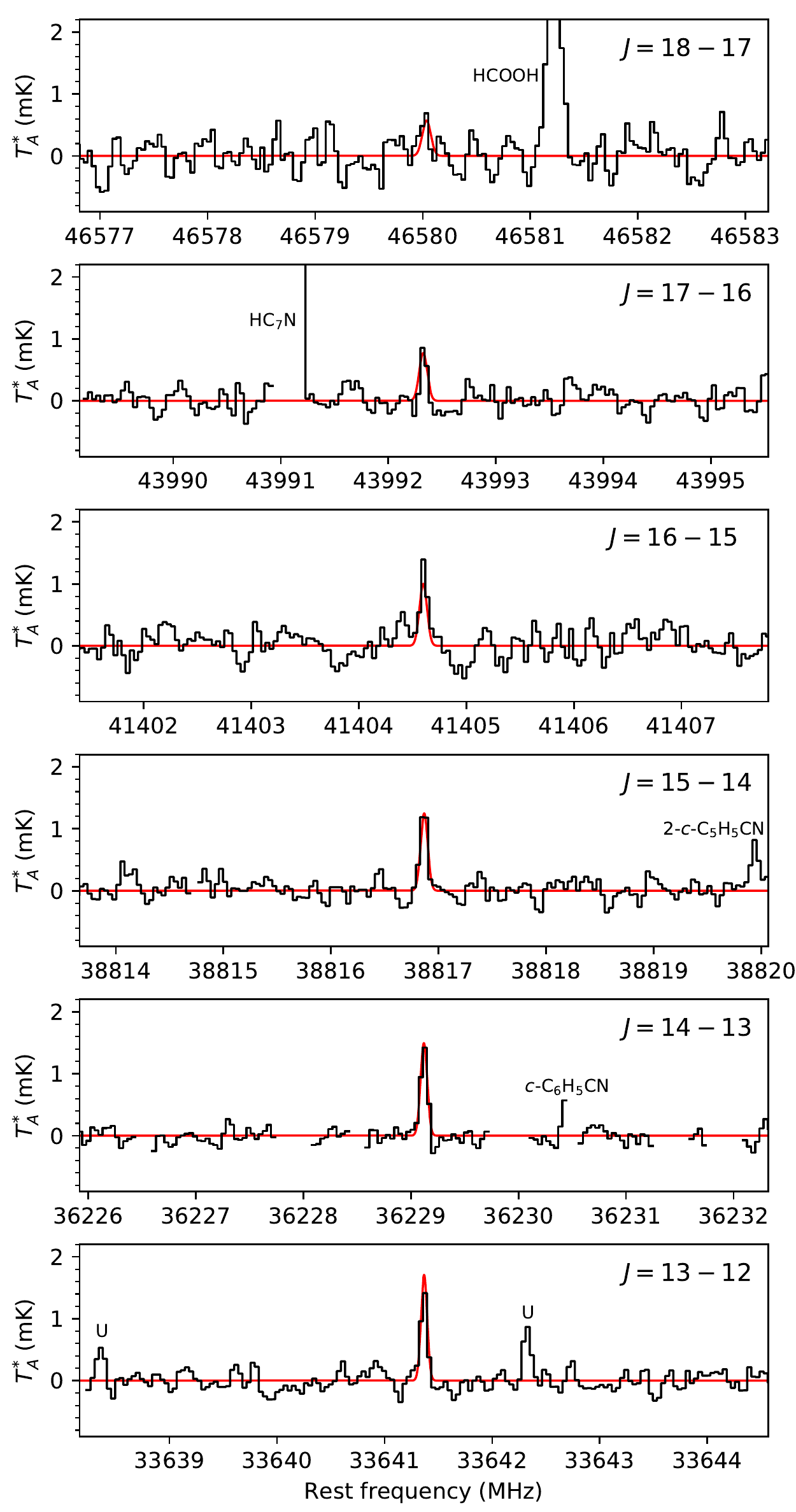}
\caption{Lines of NC$_4$NH$^+$ observed toward \mbox{TMC-1}. Blanked channels correspond to negative artifacts resulting from the frequency switching observing mode. Unidentified lines are labeled as "U". The red lines correspond to the line profiles calculated under LTE adopting a column density of 1.2\,$\times$\,10$^{10}$ cm$^{-2}$, a rotational temperature of 5.6 K, a linewidth of 0.60 km s$^{-1}$, and an emission size of 40$''$ in radius (see text).} \label{fig:lines}
\end{figure}

\begin{table*}
\small
\caption{Observed line parameters of NC$_4$NH$^+$ in \mbox{TMC-1}.}
\label{table:lines}
\centering
\begin{tabular}{cccccc}
\hline
\hline
\multicolumn{1}{c}{Transition} & \multicolumn{1}{c}{$\nu_{obs}\,^a$} & \multicolumn{1}{c}{$\nu_{obs}-\nu_{calc}$\,$^b$} & \multicolumn{1}{c}{$\Delta v$\,$^c$} & \multicolumn{1}{c}{$T_A^*$ peak} & \multicolumn{1}{c}{$\int T_A^* dv$} \\
          &  \multicolumn{1}{c}{(MHz)}     &    \multicolumn{1}{c}{(MHz)}               & \multicolumn{1}{c}{(km s$^{-1}$)}     &  \multicolumn{1}{c}{(mK)} &   \multicolumn{1}{c}{(mK km s$^{-1}$)}    \\
\hline
$J$\,=\,13-12  & 33641.365\,$\pm$\,0.010 & $-$0.002 & 0.64\,$\pm$\,0.11 & 1.48\,$\pm$\,0.13 & 1.01\,$\pm$\,0.14 \\
$J$\,=\,14-13  & 36229.121\,$\pm$\,0.010 & $+$0.000 & 0.61\,$\pm$\,0.14 & 1.49\,$\pm$\,0.11 & 0.96\,$\pm$\,0.20 \\
$J$\,=\,15-14  & 38816.863\,$\pm$\,0.010 & $-$0.003 & 0.61\,$\pm$\,0.07 & 1.39\,$\pm$\,0.13 & 0.90\,$\pm$\,0.09 \\
$J$\,=\,16-15 & 41404.606\,$\pm$\,0.010 & $+$0.006 & 0.61\,$\pm$\,0.15 & 1.33\,$\pm$\,0.20 & 0.86\,$\pm$\,0.16 \\
$J$\,=\,17-16 & 43992.330\,$\pm$\,0.010 & $+$0.006 & 0.37\,$\pm$\,0.12 & 1.00\,$\pm$\,0.16 & 0.40\,$\pm$\,0.08 \\
$J$\,=\,18-17 & 46580.010\,$\pm$\,0.020 & $-$0.025 & 0.62\,$\pm$\,0.28 & 0.64\,$\pm$\,0.25 & 0.42\,$\pm$\,0.16 \\
\hline
\end{tabular}
\tablenotea{\\
The line parameters $\nu_{\rm obs}$, $\Delta v$, $T_A^*$ peak, and $\int T_A^* dv$ as well as the associated errors were derived from a Gaussian fit to each line profile. $^a$\,Observed frequencies adopting a systemic velocity of 5.83 km s$^{-1}$ for \mbox{TMC-1} \citep{Cernicharo2020a}. $^b$\,Observed minus calculated frequencies, where calculated frequencies, $\nu_{calc}$, are computed using $B$\,=\,1293.90840 MHz and $D$\,=\,28.60 Hz (see text). $^c$\,$\Delta v$ is the full width at half maximum.
}
\end{table*}

We identified six lines in the QUIJOTE line survey whose frequencies show harmonic relations 13/14/15/16/17/18. Multiples of these relations are not possible because lines corresponding to their intermediate frequencies with similar intensities are missing. We could not assign these lines to any molecule with known rotational spectroscopy, after inspection of the private catalog of J. Cernicharo {\small MADEX}\footnote{\texttt{https://nanocosmos.iff.csic.es/?page\_id=1619}} \citep{Cernicharo2012}, the Cologne Database for Molecular Spectroscopy ({\small CDMS})\footnote{\texttt{https://cdms.astro.uni-koeln.de/}} \citep{Muller2005}, and the Jet Propulsion Laboratory ({\small JPL}) catalog\footnote{\texttt{https://spec.jpl.nasa.gov/}} \citep{Pickett1998}. The six lines are weak, with intensities around 1 mK in $T_A^*$. However, given the $T_A^*$ noise level reached by the QUIJOTE data, in the range 0.1-0.2 mK at frequencies below 45 GHz, most of the lines are detected with a high signal-to-noise ratio. The lines are shown in Fig.~\ref{fig:lines} and the frequencies measured, together with the rest of line parameters derived from a Gaussian fit, are given in Table~\ref{table:lines}.

The line frequencies can be fitted as $\nu$($J$\,$\rightarrow$\,$J$$-$$1$) = 2$B$$J$ $-$ 4$D$$J^3$, yielding a rotational constant $B$\,=\,1293.90840\,$\pm$\,0.00060 MHz and a centrifugal distortion constant $D$\,=\,28.59\,$\pm$\,1.21 Hz, with a rms of 13.5 kHz (see Table~\ref{table:param}). The six lines show a nearly perfect harmonic relation and have similar intensities. Moreover, there is no missing line that should be detected and it is not. For example, the $J$\,=\,12-11 line, predicted at 31053.604 MHz, lies outside the frequency range covered and the $J$\,=\,19-18 line, predicted at 49167.735 MHz and expected with an intensity of $T_A^*$$\sim$0.4 mK, is not detected because the $T_A^*$ noise level in this frequency range is at the level of 0.3 mK. Given these facts, it is extremely unlikely that the six lines do not arise from the same carrier.

\begin{table}
\small
\caption{Spectroscopic parameters of NC$_4$NH$^+$.}
\label{table:param}
\centering
\begin{tabular}{l@{\hspace{1.75cm}}c@{\hspace{1.75cm}}c}
\hline
\hline
Parameter & Astronomical\,$^a$ & Theoretical\,$^b$ \\
\hline
$B$ (MHz) & 1293.90840(60) & 1293.54 \\
$D$ (Hz) & 28.59(121) & 27.8 \\
rms (kHz) & 13.5 & \\
$\mu$ (D) & & 9.1 \\
\hline
\end{tabular}
\tablenoteb{\\
Numbers in parentheses are 1\,$\sigma$ uncertainties in units of the last digits. $^a$\,Values derived from a fit to the frequencies observed in \mbox{TMC-1}. The rms is the standard deviation of the fit. $^b$\,Calculated values reported in \cite{Marcelino2020}. The dipole moment is calculated at the MP2/cc-pVTZ level.
}
\end{table}

The rotational constant $B$ derived here matches very well the theoretical rotational constant calculated for protonated dicyanoacetylene (NC$_4$NH$^+$), 1293.54 MHz \citep{Marcelino2020}. This value was calculated ab initio and scaled using the experimental-to-calculated ratio of $B$ for NC$_4$N, a method that has been used previously to detect molecules such as HC$_3$O$^+$ \citep{Cernicharo2020b}, HC$_3$S$^+$ \citep{Cernicharo2021a}, CH$_3$CO$^+$ \citep{Cernicharo2021b}, H$_2$NC \citep{Cabezas2021}, HCCS$^+$ \citep{Cabezas2022a}, C$_5$H$^+$ \citep{Cernicharo2022a}, HC$_7$NH$^+$ \citep{Cabezas2022b}, and HCCNCH$^+$ \citep{Agundez2022}. In all these cases the difference between calculated and astronomical rotational constant was below 0.1\,\%, and for HC$_3$O$^+$, HC$_3$S$^+$, and CH$_3$CO$^+$ the identification was definitively confirmed by laboratory spectra. Here, the difference between the calculated rotational constant of NC$_4$NH$^+$ and the astronomical value derived from the six lines observed in \mbox{TMC-1} is 0.03\,\%. The centrifugal distortion constant calculated for NC$_4$NH$^+$ by \cite{Marcelino2020}, 27.8 Hz, is also very close to the value derived from the astronomical lines (see Table~\ref{table:param}).

Other potential candidates, apart from NC$_4$NH$^+$, are HC$_5$O$^+$ and HC$_5$NH$^+$, as discussed by \cite{Marcelino2020}. It is unlikely that HC$_5$O$^+$ is the carrier because its calculated rotational constant is significantly different, by 0.7\,\%, from that derived here. The rotational constants calculated for NC$_4$NH$^+$ and HC$_5$NH$^+$ are close, although all levels of theory predict a rotational constant 2 MHz larger for HC$_5$NH$^+$ than for NC$_4$NH$^+$, which is consistent with the assignment of the series of lines observed here to NC$_4$NH$^+$ and that observed by \cite{Marcelino2020} to HC$_5$NH$^+$. In addition, if this is the case, the differences between the calculated and the astronomical rotational constant for NC$_4$NH$^+$ and HC$_5$NH$^+$ remain small, 0.03\,\% and 0.02\,\%, respectively, while if the assignment was the reverse, the differences would be significantly larger, 0.18\,\% and 0.12\,\%, respectively. That is, we conclude that the identification of HC$_5$NH$^+$ presented by \cite{Marcelino2020} is correct and that the series of harmonically related lines presented here correspond to NC$_4$NH$^+$. The excitation of the two molecules is in line with the assignment made (see Sec.~\ref{sec:abundance}), which in any case will require of laboratory measurements for a definitive confirmation.

\subsection{Abundance of NC$_4$NH$^+$} \label{sec:abundance}

To evaluate how abundant is NC$_4$NH$^+$ in \mbox{TMC-1} we adopt a dipole moment of 9.1 D, as calculated at the MP2/cc-pVTZ level by \cite{Marcelino2020}. These authors calculated slightly higher values, in the range 9.5-9.9 D, using coupled cluster methods, although the MP2 level of theory yields probably a more accurate estimation of the dipole moment of NC$_4$NH$^+$, as suggested by the case of HC$_5$N investigated in \cite{Marcelino2020}.

\begin{figure*}
\centering
\includegraphics[angle=0,width=\columnwidth]{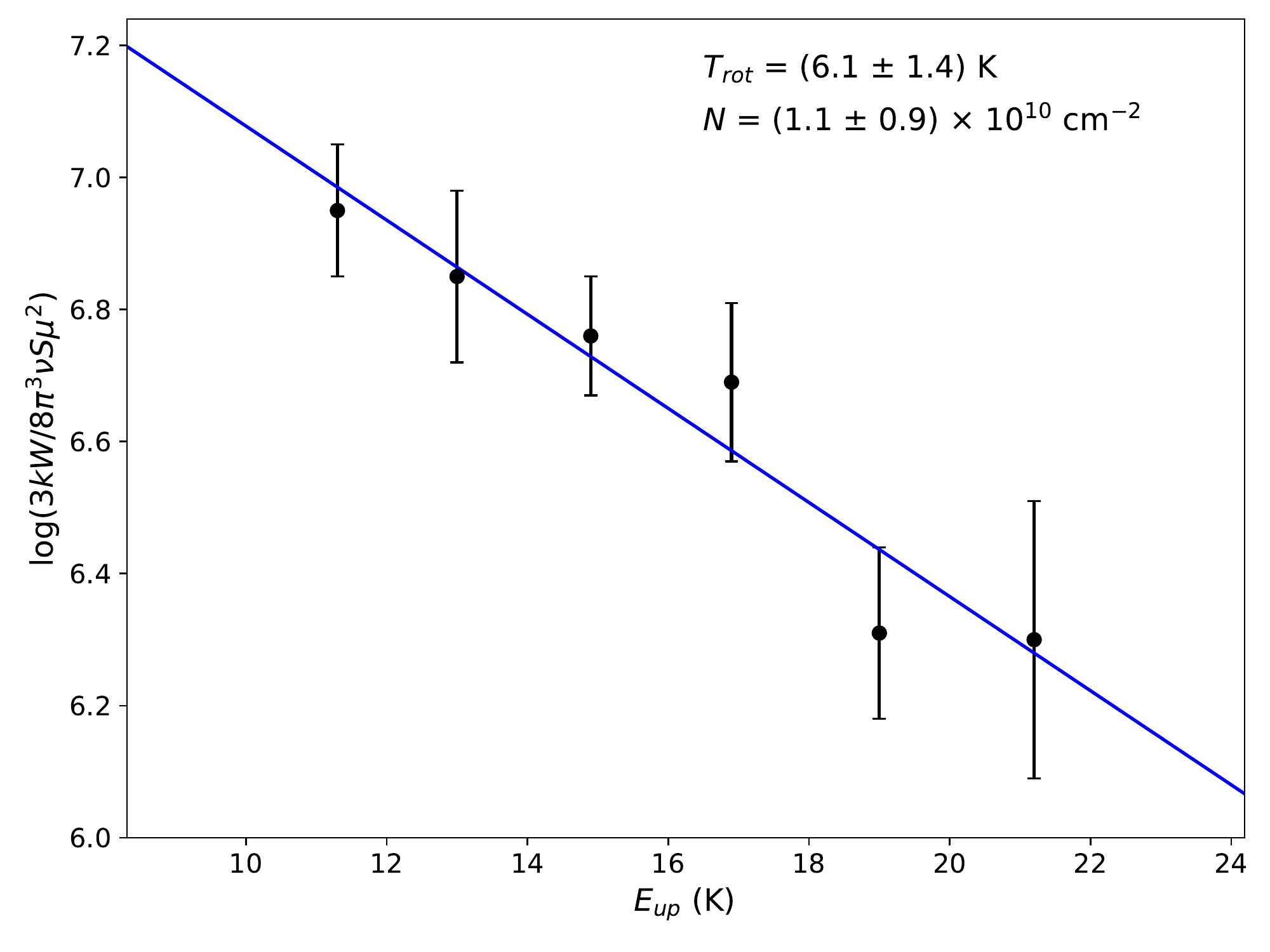} \includegraphics[angle=0,width=\columnwidth]{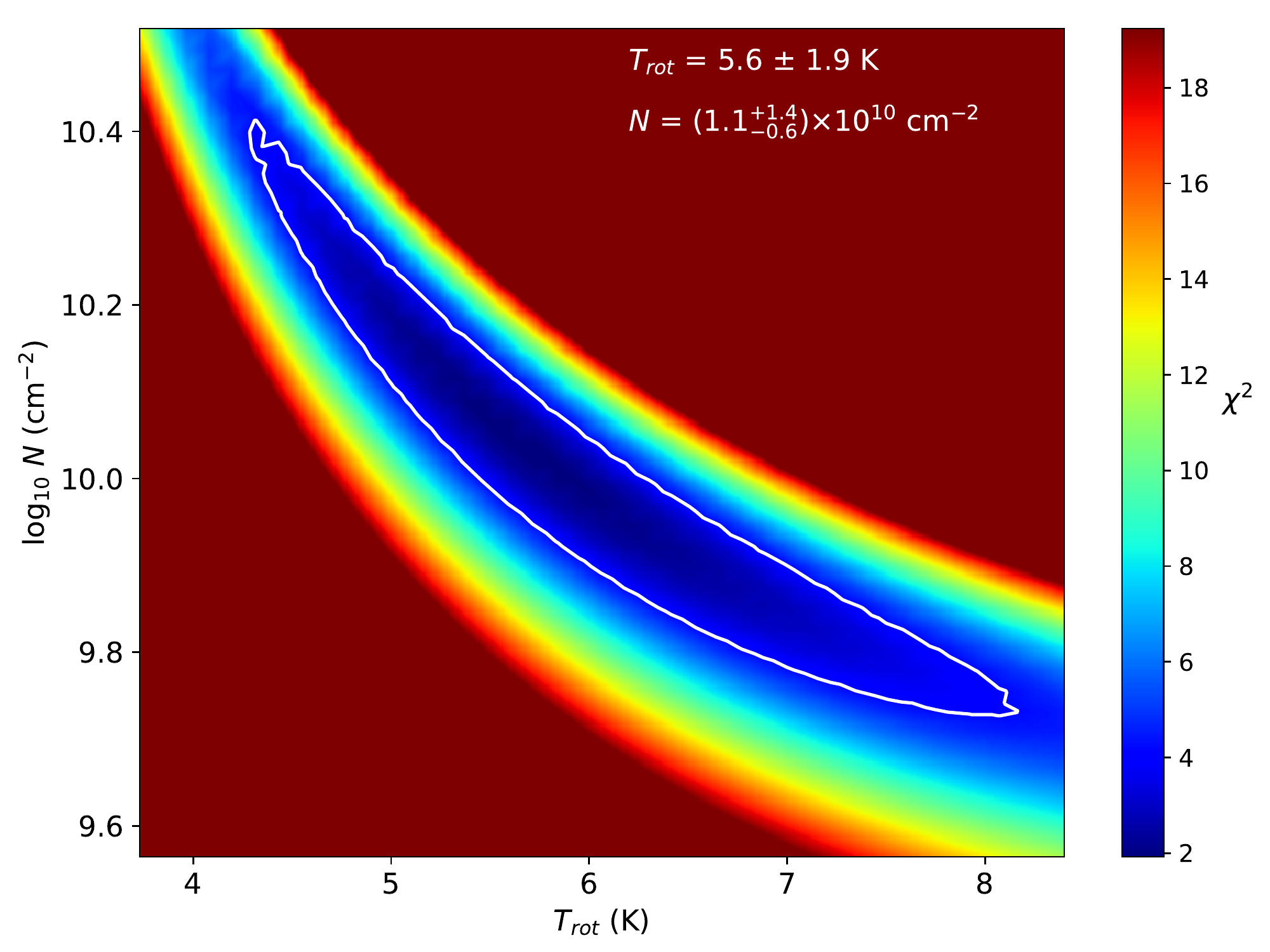}
\caption{Excitation analysis of NC$_4$NH$^+$ in \mbox{TMC-1}. In the left panel we show the rotation diagram and in the right panel we show the $\chi^2$ resulting from a LTE calculation, where the contour corresponds to the 1\,$\sigma$ level. The rotational temperature and column density derived through the two methods are very similar.} \label{fig:rtd}
\end{figure*}

Using the velocity-integrated line intensities given in Table~\ref{table:lines}, we constructed a rotation diagram, which is shown in the left panel of Fig.~\ref{fig:rtd}. We assumed that the emission of NC$_4$NH$^+$ is distributed in the sky as a circle with a diameter of 80\,$''$, as observed for several hydrocarbons in \mbox{TMC-1} \citep{Fosse2001}. We derived a rotational temperature of 6.1\,$\pm$\,1.4 K, which is in the range of the rotational temperatures derived for other molecules observed in \mbox{TMC-1}, between 5 and 10 K. The column density derived from the rotation diagram is (1.1\,$\pm$\,0.9)\,$\times$\,10$^{10}$ cm$^{-2}$. In order to get rid of some of the assumptions made by the rotation diagram method, such as the validity of the Rayleigh-Jeans limit, we carried out calculations assuming local thermodynamic equilibrium (LTE), with a single rotational temperature governing the excitation of all rotational levels, in which the rotational temperature and column density were varied. The best fit is then given by the minimum $\chi^2$, defined as $\chi^2$ = $\sum$ [($I_{calc}-I_{obs}$)/$\sigma$]$^2$, where the sum extends over the six lines, $I_{calc}$ and $I_{obs}$ are the calculated and observed velocity-integrated intensities, and $\sigma$ is the error in $I_{obs}$. In the right panel of Fig.~\ref{fig:rtd} we plot $\chi^2$ as a function of the rotational temperature and the column density. The best fit values of the rotational temperature and column density found this way are 5.6\,$\pm$\,1.9 K and (1.1\,$^{+1.4}_{-0.6}$)\,$\times$\,10$^{10}$ cm$^{-2}$, which should be more precise than the values obtained through the rotation diagram method. In any case, the differences are small and only affect the rotational temperature since the column densities obtained through the two methods are identical. The calculated line profiles using these best fit parameters are shown in Fig.~\ref{fig:lines}.

By comparing the rotational temperature derived here for NC$_4$NH$^+$, 5.6\,$\pm$\,1.9 K, with that derived for HC$_5$NH$^+$, 7.8\,$\pm$\,0.7 K \citep{Marcelino2020}, we can find an additional argument in support of the assignment to NC$_4$NH$^+$ made here. That is, the fact that the rotational temperature of NC$_4$NH$^+$ is smaller than for HC$_5$NH$^+$ is consistent with the higher dipole moment of NC$_4$NH$^+$ compared to that of HC$_5$NH$^+$, 9.1 D versus 3.6 D \citep{Marcelino2020}, because the larger the dipole moment, the larger is the critical density for thermalization, and the lower the value of the rotational temperature.

\section{Discussion} \label{sec:discussion}

\begin{figure}
\centering
\includegraphics[angle=0,width=\columnwidth]{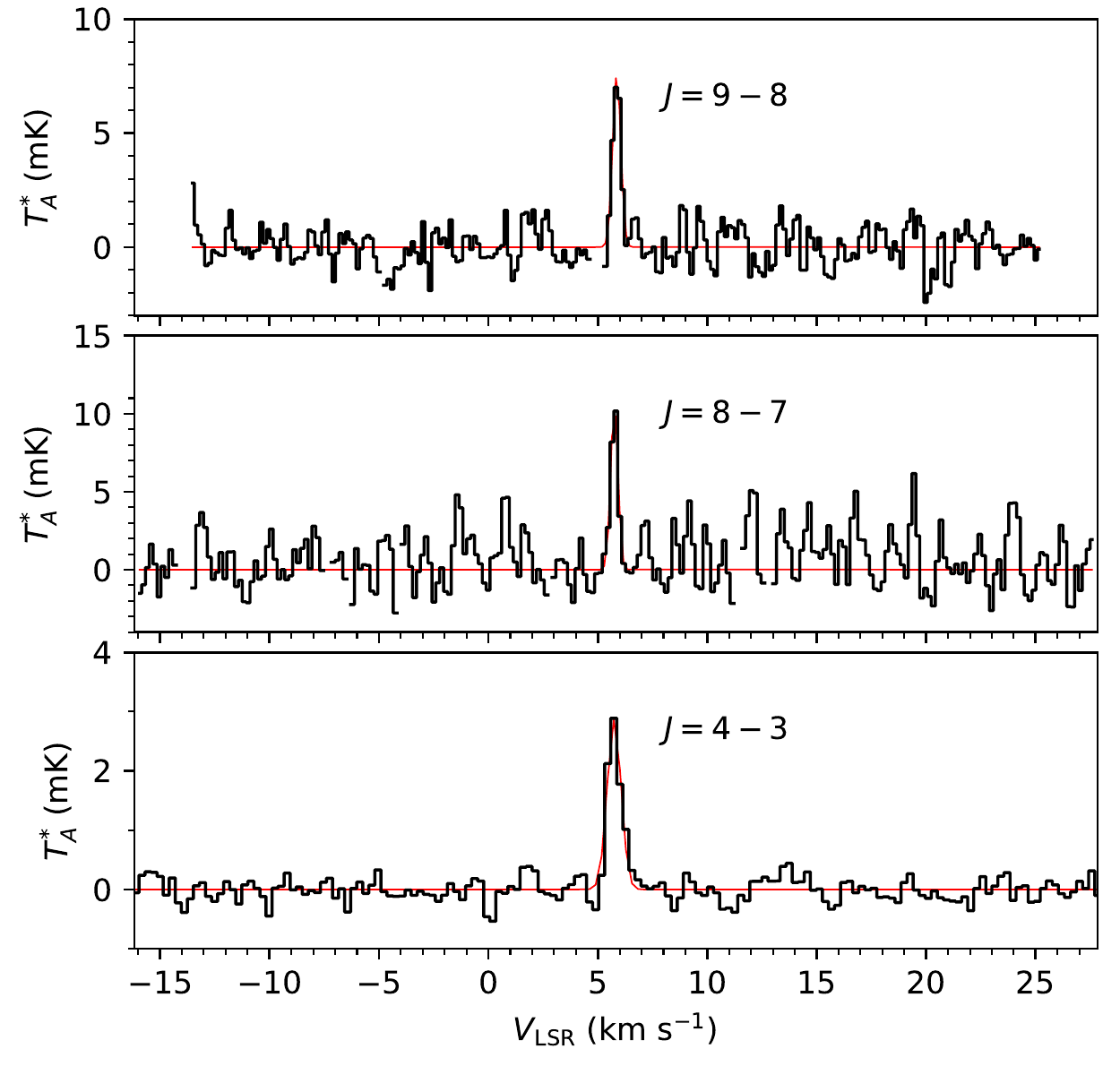}
\caption{Lines of CNCN observed toward \mbox{TMC-1}. In red we show the Gaussian fits, which yield $\int T_A^* dv$ (in mK km s$^{-1}$) of 2.27\,$\pm$\,0.13, 4.47\,$\pm$\,0.75, and 3.55\,$\pm$\,0.33 for the $J$\,=\,4-3, $J$\,=\,8-7, and $J$\,=\,9-8 lines, respectively.} \label{fig:cncn}
\end{figure}

Just as the detection of NCCNH$^+$ and CNCN \citep{Agundez2015,Agundez2018} provided indirect evidence on the existence of the highly stable, but non polar, molecule NCCN in cold dense clouds, the detection of NC$_4$NH$^+$ reported here provides indirect evidence on the presence of the non polar molecule NC$_4$N.

In the case of cyanogen (NCCN), there are two indirect proxies: the protonated form NCCNH$^+$ and the metastable isomer CNCN. The column density of NCCNH$^+$ in \mbox{TMC-1} is 8.6\,$\times$\,10$^{10}$ cm$^{-2}$ \citep{Agundez2015}, while for CNCN the detection in \mbox{TMC-1} presented by \cite{Agundez2018} was based on line stacking of four lines in the $\lambda$\,3 mm band and was therefore considered only as tentative. Here we present the confirmation of the detection of CNCN in \mbox{TMC-1} thanks to the clear detection of three lines (see Fig.~\ref{fig:cncn}): the $J$\,=\,4-3 line at 41392.912 MHz in our QUIJOTE data and the $J$\,=\,8-7 and $J$\,=\,9-8 lines at 82784.692 MHz and 93132.326 MHz, respectively, using IRAM\,30m data (observations are described in \citealt{Cabezas2022a}). We derive a rotational temperature of 10.6\,$\pm$\,1.6 K and a column density of (8.0\,$\pm$\,2.1)\,$\times$\,10$^{11}$ cm$^{-2}$, which is similar to the value of 9\,$\times$\,10$^{11}$ cm$^{-2}$ derived by \cite{Agundez2018}. Therefore, the abundance ratio CNCN/NCCNH$^+$ in \mbox{TMC-1} is $\sim$\,9.

In the case of dicyanoacetylene (NC$_4$N), the column density derived here for the protonated proxy is 1.1\,$\times$\,10$^{10}$ cm$^{-2}$. In this case there is also a metastable isomer which can be used as proxy, which is NC$_3$NC. This molecule is polar, with a calculated dipole moment of 1.11 D, and its rotational spectrum has been measured in the laboratory \citep{Huckauf1999}. At the current level of sensitivity of the QUIJOTE data, we did not detect this species, and derive a 3\,$\sigma$ upper limit to its column density of 7.3\,$\times$\,10$^{10}$ cm$^{-2}$ assuming a rotational temperature of 7.5 K, in the middle of the range 5-10 K typically found in \mbox{TMC-1}. Therefore the abundance ratio NC$_3$NC/NC$_4$NH$^+$ in \mbox{TMC-1} is $<$\,7, i.e., smaller than the ratio CNCN/NCCNH$^+$, which is $\sim$\,9. If the abundance ratios between the protonated and metastable proxies are not that different for NCCN and NC$_4$N, a deeper integration should lead to the detection of NC$_3$NC in \mbox{TMC-1}.

\begin{figure}
\centering
\includegraphics[angle=0,width=\columnwidth]{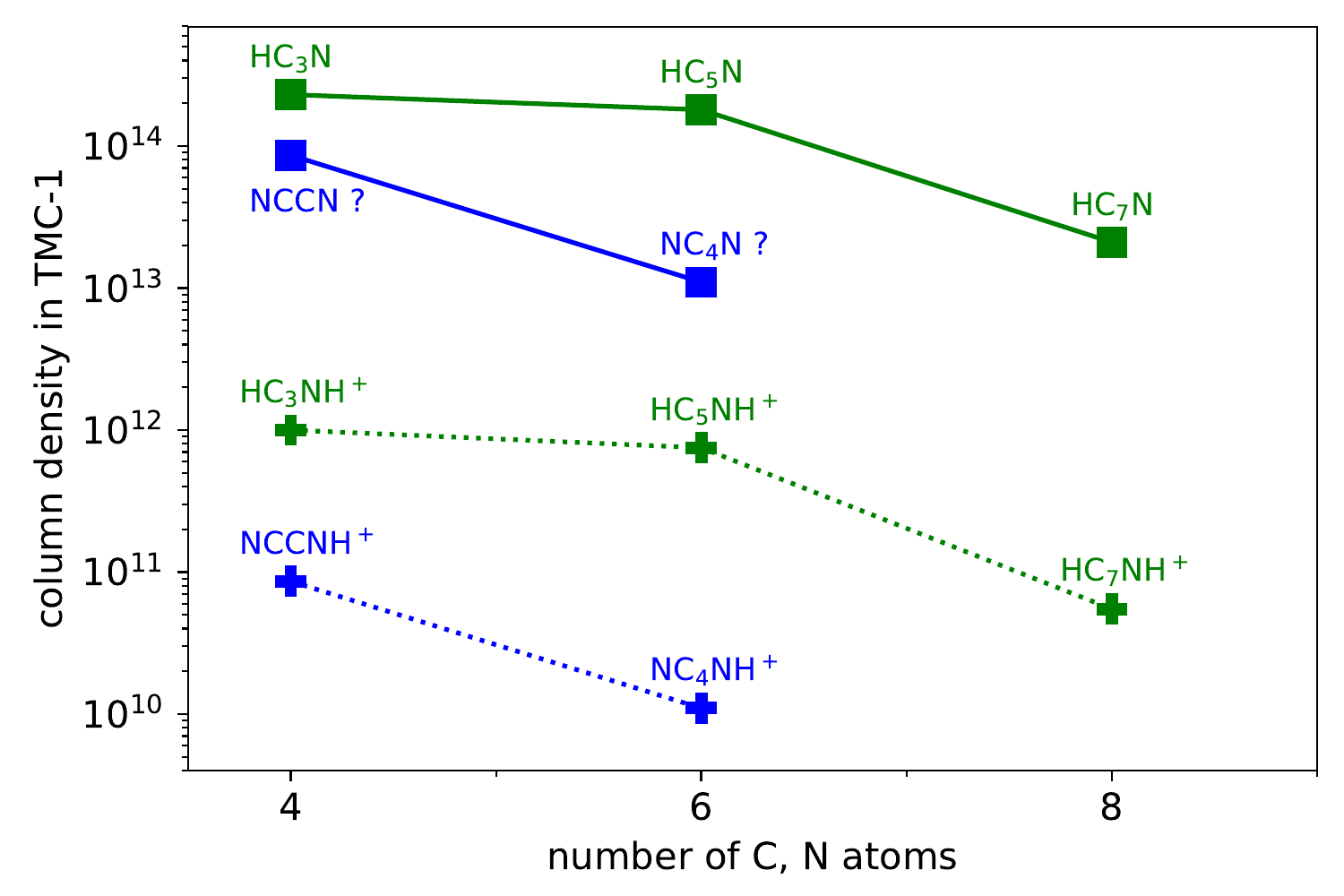}
\caption{Column densities in \mbox{TMC-1} for monocyanopolyynes (in green) and dicyanopolyynes (in blue), and their protonated forms. Observed column densities are taken from \cite{Agundez2015}, \cite{Cernicharo2020a}, \cite{Marcelino2020}, \cite{Cabezas2022b}, and this work. The column densities of NCCN and NC$_4$N are estimated from those of NCCNH$^+$ and NC$_4$NH$^+$, respectively, assuming a protonated-to-neutral abundance ratio of 10$^{-3}$ (see text).} \label{fig:cyanopolyynes}
\end{figure}

The formation of NCCNH$^+$ and NC$_4$NH$^+$ likely occurs by proton transfer to the neutral counterparts NCCN and NC$_4$N, respectively. Indeed both NCCN and NC$_4$N have high proton affinities, 674.7 kJ mol$^{-1}$ \citep{Hunter1998} and 736 kJ mol$^{-1}$ \citep{Marcelino2020}, respectively. The abundance ratio NCCNH$^+$/NCCN was calculated to be $\sim$\,10$^{-4}$ from a chemical model by \cite{Agundez2015}, although these authors noticed that the ratio may be closer to 10$^{-3}$ because the chemical model underestimated by a factor of $\sim$\,10 the protonated-to-neutral abundance ratios of related cyanides, such as HCN/HNC and HC$_3$N. From the detection of CNCN and arguments based on the expected similarity of the CNCN/NCCN and HCCNC/HCCCN ratios, \cite{Agundez2018} estimated the abundance of NCCN to be around 4.5\,$\times$\,10$^{-9}$ relative to H$_2$, which is consistent with a NCCNH$^+$/NCCN ratio of the order of 10$^{-3}$. The protonated-to-neutral abundance ratios derived from observations of cold dense clouds lie in the range 10$^{-3}$-10$^{-1}$ for neutral molecules with proton affinities larger than that of CO \citep{Agundez2022}. For the monocyanopolyynes HC$_3$N, HC$_5$N, and HC$_7$N, the protonated-to-neutral abundance ratios are a few 10$^{-3}$. If we assume a protonated-to-neutral abundance ratio of 10$^{-3}$ for NCCN and NC$_4$N, the estimated column densities of these dicyanopolyynes are somewhat lower, by factors of 3 and 15, than those of the corresponding monocyanopolyynes HC$_3$N and HC$_5$N (see Fig.~\ref{fig:cyanopolyynes}). For the next member of the series of dicyanopolyynes, NC$_6$N, we can estimate its column density to be around $\sim$\,10$^{12}$ cm$^{-2}$, assuming that the decrease in the column density with size found for HC$_5$N and HC$_7$N holds also for NC$_4$N and NC$_6$N (see Fig.~\ref{fig:cyanopolyynes}). Assuming a protonated-to-neutral ratio of 10$^{-3}$, the column density of its protonated form, NC$_6$NH$^+$, would be $\sim$\,10$^9$ cm$^{-2}$. According to the rotational constants and dipole moment calculated by \cite{Cabezas2022b}, the most intense lines of NC$_6$NH$^+$ are expected to lie in the range 10-30 GHz, although the expected brightness temperatures are as low as 0.2 mK, which would require very sensitive observations.

Now the question is how are dicyanopolyynes formed in cold dense clouds. In the case of the simplest member of the series, NCCN, its formation is thought to occur by the reaction \citep{Petrie2003}
\begin{equation}
\rm CN + HNC \rightarrow \rm NCCN + H. \label{reac:cn+hnc}
\end{equation}
Similarly, for NC$_4$N one can consider the reaction
\begin{equation}
\rm C_3N + HNC \rightarrow \rm NC_4N + H. \label{reac:c3n+hnc}
\end{equation}
If this reaction is the main source of NC$_4$N, we should expect the NC$_4$N/NCCN ratio to reflect the C$_3$N/CN ratio. In \mbox{TMC-1} the C$_3$N/CN ratio is 1.2 (Ag\'undez et al. in prep; \citealt{Pratap1997}), while the NC$_4$N/NCCN ratio can be assumed to a first approximation to be given by the NC$_4$NH$^+$/NCCNH$^+$ ratio, which is 0.13. Given that C$_3$N is as abundant as CN in \mbox{TMC-1}, one should expect NC$_4$N to be as abundant as NCCN. From the observed abundances of NCCNH$^+$ and NC$_4$NH$^+$, we infer that NC$_4$N is somewhat less abundant than NCCN, which could happen if the rate coefficient of reaction~(\ref{reac:c3n+hnc}) is lower than that of reaction~(\ref{reac:cn+hnc}) or if NC$_4$N is destroyed more rapidly than NCCN. Other plausible reactions, such as,
\begin{equation}
\rm CN + HNC_3 \rightarrow \rm NC_4N + H, \label{reac:cn+hnc3}
\end{equation}
\begin{equation}
\rm CN + HCCNC \rightarrow \rm NC_4N + H, \label{reac:cn+hccnc}
\end{equation}
are unlikely to be behind the formation of NC$_4$N in \mbox{TMC-1} because the abundance ratios HNC$_3$/HNC and HCCNC/HNC are 0.002 and 0.011, respectively \citep{Pratap1997,Cernicharo2020a}, which are much lower than the inferred NC$_4$N/NCCN ratio of 0.13.

\section{Conclusions}

We observed six harmonically related lines in our QUIJOTE line survey of \mbox{TMC-1}, which we identify as due to protonated dicyanoacetylene (NC$_4$NH$^+$). The detection of this species, together with the previous detections of NCCNH$^+$ and CNCN, support the hypothesis that dicyanopolyynes are abundant in cold dense clouds. We estimate that dicyanopolyynes are only a few times to one order of magnitude less abundant than their corresponding monocyanopolyynes.

\begin{acknowledgements}

We acknowledge funding support from Spanish Ministerio de Ciencia e Innovaci\'on through grants PID2019-106110GB-I00, PID2019-107115GB-C21, and PID2019-106235GB-I00 and from the European Research Council (ERC Grant 610256: NANOCOSMOS).

\end{acknowledgements}

\end{document}